\documentclass[doublecol,figures]{epl2}
\usepackage[english]{babel}
\usepackage{amsmath,amsthm,amsbsy}
\usepackage{amsfonts}
\usepackage{psfrag}
\usepackage{amssymb}
\usepackage{indentfirst}
\usepackage{epsfig}
\usepackage{bm}
\usepackage{wrapfig}
\usepackage{xcolor}

\title{Nonlocal effects in metallic nanoparticles: the kinetic approach outlook}
\author{P. M. Tomchuk\inst{1} \and D. V. Butenko\inst{1,2}}
\institute{
  \inst{1} Institute of Physics, National Academy of Sciences of Ukraine - 46, Nauky Avenue, Kyiv-28, 03680, Ukraine\\
  \inst{2} National University of Kyiv-Mohyla Academy - 04070 Kyiv, Ukraine}


\abstract{For the metallic nanoparticles, smaller than the free electron path, an impact of the particle's surface on the nonlocal effects emerging is shown. Light-induced current inside the particle begins to depend on the spatial derivatives of the field that leads to modification of Maxwell's equations. Consequently, the results of Mie theory as well as definitions of the dielectric function and optical conductivity should be revisited. For the sphere-shaped nanoparticle, the explicit expression for the high-frequency current with account of nonlocality is obtained. The dependence of the nonlocal contribution on the light frequency and particle's size is discussed.}

\begin{document}

\maketitle
The optical properties of metallic nanostructures became a subject of great interest last years as the methods for their fabrication \cite{VHT,KVH,WV} and experimental study (see, e.g., \cite{GdA}) had been improved. Special attention is paid to the particles with the characteristic dimensions less than $10~\mathrm{nm}$. For such small objects spatially nonlocal effects should be considered to describe an optical response correctly. This problem has been theoretically investigated in the past \cite{DF,Aga} and has received a new impetus with development of computation techniques \cite{CS,MGS,DA,FWG,LFW}.

A common approach to describe nonlocal effects in nanoobjects is based on various implementations of a hydrodynamic model \cite{RTJ,CYR,MGS,Mer,GdA2} with a dielectric function in Drude form \cite{DF,MGS,Mer,GdA2}:
\begin{equation}\label{intro}
    \epsilon(\omega,\textbf{k})=1-\frac{\omega^2_{pl}}{\omega(\omega+i\nu)-\beta^2k^2},
\end{equation}
where $\omega_{pl}$ is the plasma frequency, $\nu^{-1}$ is a relaxation time of an electron subsystem, and $\beta$ is proportional to the Fermi velocity $v_F$ and depends on the particle's dimension. 

In this Letter we calculate a nonlocal relation between the current density and the induced electric field in the framework of kinetic approach. Thus, for the particles, smaller than the free electron path, we take into account an impact of the particle's surface rigorously.

The optical properties of metallic nanoparticles (light absorption, scattering, etc.) are completely defined by the solutions of Maxwell's equations with proper boundary conditions. When the metallic nanoparticle (MN) is irradiated by plane electromagnetic wave with frequency $\omega$, the transverse components of electric and magnetic fields inside the particle satisfy equations
\begin{gather}\label{1}
   \begin{split}
    \vartriangle\!\!\textbf{E}_t+k^2\textbf{E}_t=0,\\
    \vartriangle\!\!\textbf{B}_t+k^2\textbf{B}_t=0.
   \end{split}
\end{gather}
Besides, the longitudinal component of electric field satisfies an equation
\begin{equation}\label{2}
    \nabla\textbf{E}_l=0.
\end{equation}
In Eqs. (\ref{1}) wave number $k$ is related to the complex dielectric function
\begin{gather}\label{3}
k^2=\left(\frac{\omega}{c}\right)^2\!\!\epsilon(\omega)=k_0^2\epsilon(\omega),\\
\!\!\!\!\!\!\!\!\!\text{where}\,\,\,\,\,\,\,\,\, \epsilon(\omega)=\epsilon'(\omega)+i\epsilon''(\omega)=\epsilon'(\omega)+i\frac{4\pi}{\omega}\sigma(\omega).\label{4}
\end{gather}
Here $\sigma(\omega)$ is an optical conductivity of MN.

Outside the MN incident and scattered waves satisfy Eqs. (\ref{1}) with substitution $k^2$ to $k_0^2$. Moreover, tangential components of electric and magnetic fields should be continuous at the MN's surface. The solution of Eqs. (\ref{1}) for the sphere-shaped particle yields to a well-known Mie theory.

Although the results of Mie theory are exact and general, the system (\ref{1}) itself is obtained from Maxwell's equations under the certain assumptions. In particular, the following relation between the current $\textbf{j}$, induced in the MN, and an internal electric field $\textbf{E}$ is considered
\begin{equation}\label{5}
    \textbf{j}=\sigma(\omega)\textbf{E}.
\end{equation}

The relation (\ref{5}) is justified under dipole approximation only, if the incident electric field $\textbf{E}$ can be considered spatially uniform. Furthermore, we should take into account the electron scattering on the particle's surface, when the size of the particle at least in one direction is of the one order with the electron free path \cite{TG,LPY}. In this case we have the boundary conditions not only for the fields but for the conduction electrons too. To formulate the problem rigorously, we should find a non-equilibrium addition to the Fermi distribution function, determined by an internal field $\textbf{E}$, considering this field to be unknown. Then, the current density $\textbf{j}$ can be obtained according to the formula
\begin{equation}\label{6}
    \textbf{j}(\textbf{r},t)=\frac{2e^2}{(2\pi\hbar)^3}\!\!\int\textbf{v}f(\textbf{r},\textbf{v},t)d^3(mv).
\end{equation}
Here $e$ and $m$ are the electron charge and mass respectively. Besides, the distribution function $f(\textbf{r},\textbf{v},t)$ is represented in the following way
\begin{equation}\label{7}
    f=f_0(\varepsilon)+f_1(\textbf{r},\textbf{v},t),
\end{equation}
where $f_0(\varepsilon)$ is the Fermi distribution function, and $f_1(\textbf{r},\textbf{v},t)$ is a linearized addition to the Fermi function.

Finally, we should substitute an obtained expression for the current $\textbf{j}$ into Maxwell's equations, solve them, and find an internal field $\textbf{E}$ explicitly.

So, firstly, the linearized Boltzmann equation should be solved. Assuming the time relaxation approximation, this equation takes the form
\begin{equation}\label{8}
    \frac{\partial{f_1}}{\partial{t}}+\nu f_1+\textbf{v}\frac{\partial{f_1}}{\partial{\textbf{r}}}+
    \frac{e}{m}\textbf{E}\frac{\partial{f_0}}{\partial{\textbf{v}}}=0.
\end{equation}
The function $f_1(\textbf{r},\textbf{v},t)$ should satisfy boundary conditions on the surface of MN. In the case of diffusion electron scattering these conditions take form\cite{LPY}
\begin{equation}\label{9}
    f_1(\textbf{r},\textbf{v},t)|_S=0, \,\,v_n<0,
\end{equation}
where $v_n$ is the velocity component normal to the particle surface.
The solution of Eq. (\ref{8}) with boundary conditions (\ref{9}) for the Fourier component of the distribution function $f_1$ is \cite{TT}
\begin{equation}\label{10}
    f_1(\textbf{r},\textbf{v},\omega)=-e\frac{\partial{f_0}}{\partial{\varepsilon}}\int\limits_{0}^{t_0}d\tau e^{-\tilde{\nu}(t_0-\tau)}\textbf{v}\textbf{E}[\textbf{r}-\textbf{v}(t_0-\tau),\omega],
\end{equation}
where $\tilde{\nu}\equiv\nu-i\omega$ and $t_0$ is a characteristics of Eq. (\ref{8})\cite{TT}
\begin{equation}\label{11}
    t_0=\frac{1}{v^2}\left[\textbf{r}\textbf{v}+
    \sqrt{(R^2-\textbf{r}^2)\textbf{v}^2+(\textbf{r}\textbf{v})^2}\right],
\end{equation}
here $R$ is a radius of a spherical particle.

The solution (\ref{10}) is obtained for the sphere-shaped MN. This result can be easily extended to the ellipsoid-shaped particles by deformation of the coordinate system\cite{GT}.

The expression (\ref{10}) takes into account a spatial dispersion and transforms into well-known result when the particle is large ($R\rightarrow\infty$). Actually, let us consider a plane wave in Eq. (\ref{8}):
\begin{equation}\label{12}
    \textbf{E}(\textbf{r},t)=\textbf{E}_0e^{i(\textbf{kr}-\omega t)}.
\end{equation}
Then, from the solution (\ref{10}) we obtain:
\begin{equation}\label{13}
    f_1(\textbf{k},\omega)=-e\textbf{E}\textbf{v}\frac{\partial{f_0}}{\partial{\varepsilon}}
    \frac{1-\exp\{-t_0[\tilde{\nu}-i(\omega-\textbf{kv})]\}}{\tilde{\nu}-i(\omega-\textbf{kv})}.
\end{equation}
Now, using expressions (\ref{5}), (\ref{6}), and (\ref{13}) we find a tensor of optical conductivity
\begin{equation}\label{14}
\begin{split}
\sigma_{\alpha\beta}(\textbf{k},\omega)&=-\frac{2e^2}{(2\pi\hbar)^3}\!\int d^3v\frac{1-\exp\{-t_0[\tilde{\nu}-i(\omega-\textbf{kv})]\}}{\tilde{\nu}-i(\omega-\textbf{kv})}\times\\
&\times v_\alpha v_\beta\frac{\partial{f_0}}{\partial{\varepsilon}}.
\end{split}
\end{equation}
The result (\ref{14}) takes the standard form for the tensor of optical conductivity accounting a spatial dispersion, when $R\rightarrow\infty$:
\begin{equation}\label{15}
\sigma_{\alpha\beta}(\textbf{k},\omega)=-\frac{2e^2}{(2\pi\hbar)^3}\!\int d^3v\frac{v_\alpha v_\beta\frac{\partial{f_0}}{\partial{\varepsilon}}}{\tilde{\nu}-i(\omega-\textbf{kv})}.
\end{equation}
However, we can not set an internal electric field in the form (\ref{12}). In this form we can only set an incident field, but internal and scattered fields should be obtained from the Maxwell's equations' solutions.

If the wave vector of an incident wave satisfies the condition $kR\ll1$, an induced electric field inside the spherical (or ellipsoidal) MN can be considered spatially uniform. When the parameter $kR$ increases, an internal field $\textbf{E}(\textbf{r},\omega)$ begins to depend on the coordinates inside the MN. As we see from Eq. (\ref{10}), the distribution function $f_1(\textbf{r},\textbf{v},\omega)$ depends on all of the values of an internal electric field $\textbf{E}[\textbf{r}-\textbf{v}(t_0-\tau),\omega]$ in the interval $0\leq\tau\leq t_0$. This causes the nonlocality between the electric field and induced current. If the nonlocality is not strong, that depends on the value of parameter $kR$, the electric field inside the MN can be expanded in the Taylor series near $\textbf{E}(\textbf{r},\omega)$. In this case from Eq. (\ref{10}) we obtain
\begin{equation}\label{16}
\begin{split}
    &f_1(\textbf{r},\textbf{v},\omega)=-e\frac{\partial{f_0}}{\partial{\varepsilon}}\left\{\textbf{v}\textbf{E}(\textbf{r},\omega)+
    \left(\textbf{v}\frac{\partial{}}{\partial{\textbf{r}}}\right)[\textbf{v}\textbf{E}(\textbf{r},\omega)]\frac{\partial{}}{\partial{\tilde{\nu}}}+\right.\\
    &\left.+\frac{1}{2}\left(\textbf{v}\frac{\partial{}}{\partial{\textbf{r}}}\right)^2[\textbf{v}\textbf{E}(\textbf{r},\omega)]
    \frac{\partial{}^2}{\partial{\tilde{\nu}}^2}\right\}\frac{1-e^{-\tilde{\nu}t_0}}{\tilde{\nu}}.
\end{split}
\end{equation}
Now, using formulas (\ref{6}) and (\ref{16}) we can find an expression for the current density $\textbf{j}$. The characteristics $t_0$ depends on the angle between vectors $\textbf{r}$ and $\textbf{v}$. That is why it is convenient to choose a direction of the vector $\textbf{r}$ as an axis $Oz$. This procedure will be done in two steps. Firstly, we represent an operator $\textbf{v}\frac{\partial{}}{\partial{\textbf{r}}}$ in a laboratory coordinate system where the direction of $\textbf{E}(\textbf{r},\omega)$ is set as an axis $Oz$. Therefore,
\begin{equation}\label{17}
    \textbf{v}\frac{\partial{}}{\partial{\textbf{r}}}=v_x^{(0)}\frac{\partial{}}{\partial{x}}+v_y^{(0)}\frac{\partial{}}{\partial{y}}+
    v_z^{(0)}\frac{\partial{}}{\partial{z}}.
\end{equation}
Then, we rotate the system $v_x^{(0)},v_y^{(0)},v_z^{(0)}$ in such a way that the axis $Oz$ coincides with the direction of the vector $\textbf{r}$. As a result, we obtain
\begin{equation}\label{18}
    \textbf{v}\frac{\partial{}}{\partial{\textbf{r}}}=v_x\frac{1}{r}\frac{\partial{}}{\partial{\xi}}
    +v_y\frac{1}{r\sin\xi}\frac{\partial{}}{\partial{\eta}}+v_z\frac{\partial{}}{\partial{r}}.
\end{equation}
Here $v_x,v_y,v_z$ is a velocity components in the new coordinate system, and angles $\xi$ and $\eta$ define the direction of the vector $\textbf{r}$ in the laboratory coordinate system.

Now, substituting an expansion (\ref{16}) and an operator (\ref{18}) into the formula (\ref{6}), we obtain an expression for a component of the current density along the electric field $\textbf{E}(\textbf{r},\omega)$
\begin{equation}\label{19}
    \begin{split}
    &j_z(\textbf{r},\omega)\!=\!\frac{3ne^2}{4m}\!\left\{E\!\!\int\limits_{0}^{\pi}\!\! d\theta\sin\theta\left(\sin^2\!\xi\sin^2\!\theta\!+\!2\cos^2\!\xi\cos^2\!\theta\right)\!+\right.\\
    &+\!v_F\!\!\!\int\limits_{0}^{\pi}\!\!d\theta\sin\theta\cos\theta
    \left[\frac{\partial{E}}{\partial{r}}\left(\sin^2\!\xi\sin^2\!\theta+2\cos^2\!\xi\cos^2\!\theta\right)-\right.\\
    &\left.\left.-\frac{1}{r}\frac{\partial{E}}{\partial{\xi}}
    \sin2\xi\sin^2\!\theta\right]\!\frac{\partial{}}{\partial{\tilde{\nu}}}\right\}\!\frac{1-e^{-\tilde{\nu}t_0}}{\tilde{\nu}}.
    \end{split}
\end{equation}
Here $\theta$ is an angle between vectors $\textbf{r}$ and $\textbf{v}$, and $E$ is a magnitude of an electric field $\textbf{E}(\textbf{r},\omega)$. Calculating expression (\ref{19}) we assume that $kT\ll\varepsilon_F$, so that $\partial{f_0}/\partial{\varepsilon}\approx\delta(\varepsilon-\varepsilon_F)$ (where $\varepsilon_F$ being the Fermi energy). Moreover, we introduce the density of conduction electrons in terms of the Fermi velocity
\begin{equation}\label{1a}
    n=\frac{8\pi}{3}\left(\frac{mv_F}{2\pi\hbar}\right)^3.
\end{equation}

\begin{figure}[h!]
\begin{center}
 \begin{minipage}[h]{1\linewidth}
 \includegraphics[width=1\linewidth]{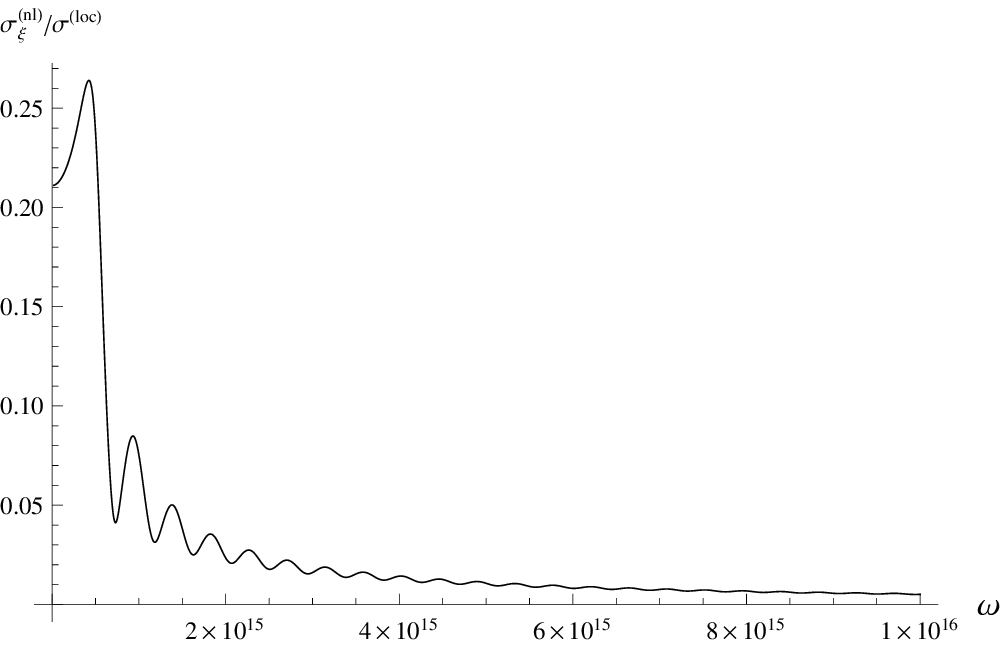}\\ a)
 \end{minipage}
 \hfill
 \begin{minipage}[h]{1\linewidth}
 \includegraphics[width=1\linewidth]{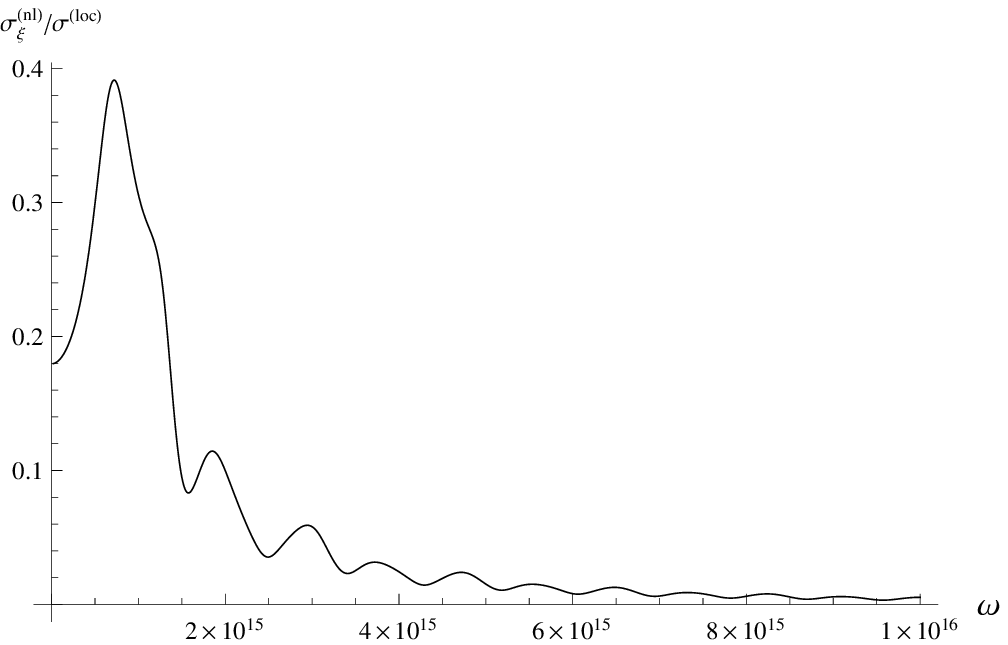}\\ b)
 \end{minipage}
 \hfill
 \begin{minipage}[h]{1\linewidth}
 \includegraphics[width=1\linewidth]{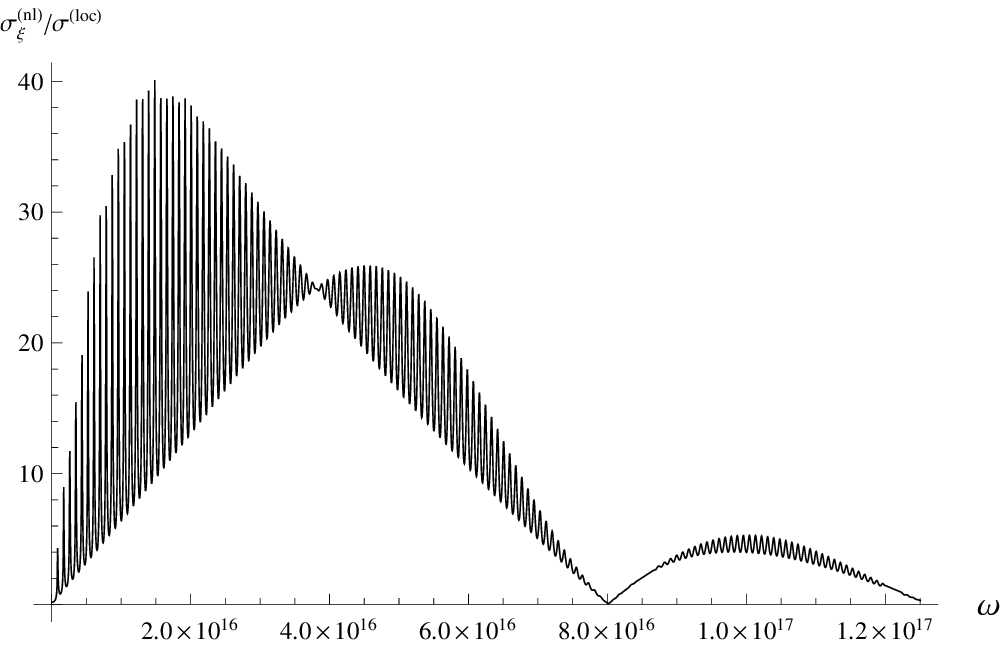}\\ c)
 \end{minipage}
 \caption{The ratio of the "nonlocal" and "local" coefficients $\sigma^{(nl)}_{\xi}/\sigma^{(loc)}$ as a function of the frequency $\omega$ at the angle $\xi~=~\pi/4$ and for different values of $r$: a) $r=R$, b) $r=0.5R$, c) $r=0.01R$. All the plots are performed for the sphere-shaped Au nanoparticles with the radius $R=10\,\mathrm{nm}$.}\label{fig1}
 \end{center}
\end{figure}

Obviously, the nonlocal term in formula (\ref{19}) will sufficiently modify Maxwell's equations only if its value is comparable with the local term. Although, we can not compare them accurately if an internal field $\textbf{E}(\textbf{r},\omega)$ is unknown, but we can estimate the ratio of these terms under the certain reasonable assumptions. Firstly, let us rewrite the expression (\ref{19}) in the next form:
\begin{equation}\label{2a}
    j_z(\textbf{r},\omega)=\sigma^{(loc)}E+\sigma^{(nl)}_r\frac{\partial{E}}{\partial{r}}-\sigma^{(nl)}_{\xi}\frac{\partial{E}}{\partial{\xi}},
\end{equation}
where the coefficients $\sigma^{(loc)}$, $\sigma^{(nl)}_r$, and $\sigma^{(nl)}_{\xi}$ are defined via such expressions:
\begin{gather}
    \sigma^{(loc)}\!\!=\!\!\frac{3ne^2}{4m}\!\!\!\int\limits_{0}^{\pi}\!\!\! d\theta\sin\!\theta\!\left(\sin^2\!\xi\!\sin^2\!\theta\!+\!2\!\cos^2\!\xi\!\cos^2\!\theta\right)\!\frac{\partial{}}{\partial{\tilde{\nu}}}\!\frac{1\!-\!e^{-\tilde{\nu}t_0}}{\tilde{\nu}},\label{3a}\\   \begin{split}\label{4a}
    \sigma^{(nl)}_r\!&=\!\frac{3ne^2 v_F}{4m}\!\!\!\int\limits_{0}^{\pi}\!\!\!d\theta\sin\!\theta\cos\!\theta\!
    \left(\sin^2\!\xi\sin^2\!\theta\!+\!2\cos^2\!\xi\cos^2\!\theta\right)\times\\
    &\times\frac{\partial{}}{\partial{\tilde{\nu}}}\!\frac{1\!-\!e^{-\tilde{\nu}t_0}}{\tilde{\nu}},
    \end{split}\\
    \!\!\!\!\!\!\!\!\!\!\!\!\!\!\sigma^{(nl)}_{\xi}=\frac{3ne^2 v_F}{4m}\frac{\sin\!2\xi}{r}\!\!\!\int\limits_{0}^{\pi}\!\!d\theta\sin^3\!\theta\cos\!\theta
    \frac{\partial{}}{\partial{\tilde{\nu}}}\!\frac{1\!-\!e^{-\tilde{\nu}t_0}}{\tilde{\nu}}.\label{5a}
\end{gather}
Varying the angle $\xi$, we can always make one of the coefficients $\sigma^{(nl)}_r$ or $\sigma^{(nl)}_{\xi}$ to be an infinitesimal value. That's why we can compare one of the nonlocal terms with the local one to make some conclusions about the contribution of the nonlocal part. The coefficients $\sigma^{(loc)}$ and $\sigma^{(nl)}_{\xi}$ are of the same dimension, so it is natural to compare their values. This comparison will give us an adequate estimation of the nonlocal term's contribution if a value of the derivative $\partial{E}/\partial{\theta}$ is comparable with a value of the field $E$ itself. This assumption seems to be reasonable, so we present in Fig.~\ref{fig1} the aforementioned ratio as a function of a frequency $\omega$.

Here we need to note that coefficients $\sigma^{(nl)}_{\xi}$ and $\sigma^{(loc)}$ are complex values, that's why an ordinate in Fig.~\ref{fig1} is a ratio of their absolute values.

As we see from Fig.~\ref{fig1}, the contribution of the nonlocal term to the current density can be significant enough all over the particle. Furthermore, frequency dependence of the "nonlocal/local" ratio varies with the distance from the point where the current is observed to the center of the particle. If the point is far from the center, the maximum of the nonlocal contribution is in infrared zone (near $5\!\cdot\!10^{14} \,\mathrm{s}^{-1}$ at the surface), and then the function $\sigma^{(nl)}_{\xi}\!/\sigma^{(loc)}(\omega)$ demonstrates oscillating behavior with rapid damping when frequency increases. As the distance to the center decreases, the maximum grows and shifts to the higher frequencies. The Fig.~1c) shows that a contribution of the nonlocal part near the particle's center is significant even at the frequencies much higher than the resonance frequency for the Au sphere ($\Omega_{sph}=\omega_{pl}/\sqrt{3}\approx7.9\!\cdot\!10^{15}\,\mathrm{s}^{-1}$).

Considering an enhancement of the nonlocal effect near the center of MN, it is interesting to analyze the expression (\ref{19}) for small values of the distance $r$. According to the formula (\ref{11}), when $r\rightarrow0$, we obtain an expansion
\begin{equation}\label{20}
    \begin{split}
    e^{-\tilde{\nu}t_0}\!\approx\!e^{-\tilde{\nu}/2\nu_S}\!\!\left[1\!-\!\frac{\tilde{\nu}\,r\cos\theta}{2\nu_SR}\!+\!
    \frac{\tilde{\nu}\,r^2}{4\nu_SR^2}\!\left(\!\frac{\tilde{\nu}}{2\nu_S}\!\cos^2\!\theta\!+\!\sin^2\!\theta\!\right)\!\right]\!\!.
    \end{split}
\end{equation}
After substituting (\ref{20}) into (\ref{19}) and fulfilling the integrals, we obtain the following expression for the current density
\begin{equation}\label{21}
    \begin{split}
    &j_z(\textbf{r},\omega)\!=\!\frac{ne^2}{m\tilde{\nu}}\!\left\{1\!-\!e^{-\tilde{\nu}/2\nu_S}\!-\!
    \frac{\tilde{\nu}\,r^2}{20\nu_SR^2}e^{-\tilde{\nu}/2\nu_S}\times\right.\\
    &\left.\times\left[3\!+\!\frac{\tilde{\nu}}{\nu_S}\!+\!\left(\frac{\tilde{\nu}}{2\nu_S}\!-\!1\right)\cos2\xi\right]\right\}E-\\
    &-\frac{ne^2}{10m\nu_S}e^{-\tilde{\nu}/2\nu_S}\left\{(2+\cos2\xi)r\frac{\partial{E}}{\partial{r}}-\sin2\xi\frac{\partial{E}}{\partial{\xi}}\right\}.
    \end{split}
\end{equation}
Here we introduce the parameter $\nu_S=v_F/2R$, that has meaning of electron-surface collisions' frequency. If this "surface" frequency is much less than the "bulk" frequency $\nu$, we obtain a well-known result from the formula (\ref{21})
\begin{equation}\label{22}
    j_z\approx\frac{ne^2}{m(\nu-i\omega)}E, \,\,\,\,\,\,\,\nu_S\ll\nu.
\end{equation}
In the opposite case, when the electron scattering at the MN's surface is dominant, an expression (\ref{21}) takes the form:
\begin{equation}\label{23}
    j_z\!\approx\!\frac{ne^2}{2m\nu_S}\!\!\left\{\!E\!-\!\frac{1}{5}\!\!
    \left[\!(2\!+\!\cos\!2\xi)r\frac{\partial{E}}{\partial{r}}\!-\!\sin\!2\xi\frac{\partial{E}}{\partial{\xi}}\right]\!\right\},
    \,\,\,\,\nu_S\gg\nu.
\end{equation}
As we see from these asymptotic formulas, the nonlocal effect vanishes when the MN's size is much larger than the free electron path. If the inverse relation is true, a contribution of the local and nonlocal parts to the induced current are of the same order of magnitude. It means that an impact of the particle's surface into the electron scattering causes the nonlocal effects inside the MN.

So, when the nonlocality is strong, the dependence of the induced current $\textbf{j}(\textbf{r},t)$ on the derivatives $\partial{E}/\partial{r}$ and $\partial{E}/\partial{\xi}$ leads to a noticeable modification of Maxwell's equations' solutions. Consequently, results of Mie theory as well as a definition of the dielectric function, conductivity, and related quantities should be reconsidered.


\begin{thebibliography}{99}

\bibitem[1]{VHT}    \Name{Van Duyne R. P., Hulteen J. C. \and Treichel D. A.}
  \REVIEW{J. Chem. Phys.}{99}{1993}{2101}.

\bibitem[2]{KVH}    \Name{Kahl M.,Voges E. \and Hill W.}
  \REVIEW{Spectrosc. Eur.}{10}{1998}{12}.

\bibitem[3]{WV}    \Name{Willets K. A. \and Van Duyne R. P.}
  \REVIEW{Annu. Rev. Phys. Chem.}{58}{2007}{267}.

\bibitem[4]{GdA}    \Name{Garcia de Abajo F. J.}
  \REVIEW{Rev. Mod. Phys.}{82}{2010}{209}.

\bibitem[5]{DF}    \Name{Dasgupta B. B. \and Fuchs R.}
  \REVIEW{Phys. Rev. B}{24}{1981}{554}.

\bibitem[6]{Aga}    \Name{Agarwal G. et al.}
  \REVIEW{Phys. Rev. B}{10}{1974}{1447}.

\bibitem[7]{CS}    \Name{Coronado E. A.  \and Schatz G. C.}
  \REVIEW{J. Chem. Phys.}{119}{2003}{3926}.

\bibitem[18]{CYR}  \Name{Christensen T., Yan W., Raza S., Jauho A.-P., Asger Mortensen N. \and Wubs M.}
  \REVIEW{ACS Nano}{8}{2014}{1745}.

\bibitem[19]{RTJ}  \Name{Raza S., Toscano G., Jauho A.-P., Wubs M. \and Asger Mortensen N.}
  \REVIEW{Phys. Rev. B}{84}{2011}{121412}.

\bibitem[8]{MGS}     \Name{McMahon J. M., Gray S. K. \and Schatz G. C.}
  \REVIEW{Phys. Rev. B}{82}{2010}{035423}.

\bibitem[9]{DA}    \Name{David C. \and Garcia de Abajo F. J.}
  \REVIEW{J. Phys. Chem. C}{115}{2011}{19470}.

\bibitem[10]{FWG}    \Name{Fernandez-Dominguez A. I., Wiener A., Garcia-Vidal F. J., Maier S. A. \and Pendry J. B.}
  \REVIEW{Phys. Rev. Lett.}{108}{2012}{106802}.

\bibitem[11]{LFW}    \Name{Luo Y., Fernandez-Dominguez A. I., Wiener A., Maier S. A. \and Pendry J. B.}
  \REVIEW{Phys. Rev. Lett.}{111}{2013}{093901}.

\bibitem[12]{Mer}   \Name{Mermin N. D.}
  \REVIEW{Phys. Rev. B}{1}{1970}{2362}.

\bibitem[13]{GdA2}   \Name{Garcia de Abajo F. J.}
  \REVIEW{J. Phys. Chem. C}{112}{2008}{17983}.

\bibitem[14]{TG}    \Name{Tomchuk P. M. \and Grigorchuk N. I.}
  \REVIEW{Phys. Rev. B}{73}{2006}{155423}.

\bibitem[15]{LPY}    \Name{Lesskis A. G., Pasternak V. E. \and Yushkanov A. A.}
  \REVIEW{JETP}{83}{1982}{310}.

\bibitem[16]{TT}    \Name{Tomchuk P.M. \and Tomchuk B. P.}
  \REVIEW{JETP}{112}{1997}{661}.

\bibitem[17]{GT}    \Name{Grigorchuk N. I. \and Tomchuk P. M.}
  \REVIEW{Phys. Rev. B}{84}{2011}{085448}.

\end{thebibliography}
\end{document}